\documentclass[12pt]{article}
\usepackage{amssymb,amsmath}
\begin{document}

%%%%%%%%%%%%%%%%%%%%%%%%%%%%%%%%%%%%%%%%%%%%%%%%%%%%%%%%%%%%%%%%%%%%%
%%%%%%%%%%%%%%      TITLE PAGE     %%%%%%%%%%%%%%%%%%%%%%%%%%%%%%%%%%
%%%%%%%%%%%%%%%%%%%%%%%%%%%%%%%%%%%%%%%%%%%%%%%%%%%%%%%%%%%%%%%%%%%%%

\sloppy
\title
%{\hfill{\normalsize\sf FIAN/TD/01-15}    \\
 %           \vspace{1cm}
{\Large  Classical and quantum scattering by a gravitational center }

\author
 {
       { A.I.Nikishov}
          \thanks
             {E-mail: nikishov@lpi.ru}
  \\
               {\small \phantom{uuu}}
  \\
 }
%
%--------------------------------------------------------------------
\maketitle
%--------------------------------------------------------------------
%%%%%%%%%%%%%%%%%%%%%%%%%%%%%%%%%%%%%%%%%%%%%%%%%%%%%%
\begin{abstract}
The small angle  scattering (by a gravitational field)  of classical and quantum particles is considered and compared. It is suggested that the differences in small angle scattering of particles with spin 0, 1, 2  are due to the nonzero probability of forward scattering for
particles described by a wave packet. It is suggested that measurements of the deflection of light in the vicinity of the Sun will decide which coordinate system is the privileged one.
\end{abstract}

\section{Introduction}
When a photon is deflected by the Sun, we can know with good accuracy its
impact parameter $\rho$, its frequency and the coordinates of place where it is observed.
We note that we always observe at a finite distance from the Sun and in principle the
place of observation is at our disposal. Classical theory permits to predict the results
of such observations.
So we expect that the classical approach should be applicable in rather wide range of impact
parameters and  values of $r_{g}=2Gm/c^2$. The classical cross section may also serve as a guide for quantum  calculation beyond the Born approximation. To some extent this
 should be also true in Coulomb scattering [1].

For a scattering with the momentum transfer $q$ the impact parameter $\rho$
is of order of $\hbar/q$ and the formation length of this process is of order
of several $\hbar/q$. Beyond this length the particle is unable to obtain the
momentum transfer $q$. In quantum theory in Born approximation the particle
obtains the required momentum transfer by interacting with only one graviton.
In principle the quantum particle can pass the formation length without
interaction or interacting with several gravitons in such a way that no momentum
transfer is passed to it. It is very interesting to know the probability for this process.
If it is nonzero, then the concept of curved space time is of only limited
validity. Due to the possibility of flying by without deflection the
quantum cross section should be smaller then the classical one. So the detailed
study of classical and quantum scattering is of interest in many respects.

\section{Small angle classical scattering by the Schwarzschild field}
{\bf A) Particle with mass.}\\
The trajectories in the Schwarzschild field were studied in a number of papers
[2-4], see also [5].
If the standard coordinate system
$$
ds^2=\left(1-\frac{r_g}{r}\right)(dx^0)^2-r^2(\sin^2\theta d\varphi^2+d\theta^2)-
{\left(1-\frac{r_g}{r}\right)}^{-1}dr^2                                       \eqno(1)
$$
is used, the equation governing the scattering trajectory has the simplest form
 $$
  \frac{du}{d\vartheta}=\pm\sqrt{f(u)}, \quad u=\frac{\rho}{r},
  \quad \delta=\frac{r_g}{\rho},\quad r_g=\frac{2GM}{c^2},
$$
$$
  f(u)=1-u^2+\varkappa u\delta+u^3\delta=\delta(u-u_1)(u-u_2)(u-u_3),\quad
  \varkappa=\beta^{-2}-1.                                                        \eqno(2)
 $$

 Here $\beta $ is velocity at infinity in units of $c$,
 the signs $+$ and $-$ in front of the square root refer to the first and the
 second halves of trajectory respectively. From (2) the half angle between the
  asymptotes is
  $$
  \vartheta_{1/2}=\int_0^{\vartheta_{1/2}}d\vartheta=\int_0^{u_2}
  \frac{du}{\sqrt{f(u)}}.                                                        \eqno(3)
  $$
This is the contribution from the first part of the trajectory.
  It is assumed that $u_1<u_2<u_3$.
  Similarly, for the second half:
  $$
 2\vartheta_{1/2}-\vartheta_{1/2} =\int_{\vartheta_{1/2}}^{2\vartheta_{1/2}}d\vartheta=
  -\int_{u_2}^0\frac{du}{\sqrt{f(u)}}=\int_{0}^{u_2}\frac{du}{\sqrt{f(u)}}=
 \vartheta_{1/2}.                                                                  \eqno(4)
  $$
  So the angle between asymptotes is $2\vartheta_{1/2}$. The scattering angle is
$\theta=2\vartheta_{1/2}-\pi.$ Each half of the trajectory contributes $\vartheta_{1/2}.$

  Now the expressions for roots $u_1, u_2, u_3$ as functions of $\delta$
  and $\varkappa$ can
  be obtained by the method of Newton: if $x^{(0)}$ is the zero order
  approximation for the root of $f(x)=0$, then in the first approximation we
  have $x^{(1)}= x^{(0)}-\frac{f(x^{(0)})}{f'(x^{(0)})}$. In the second
  approximation $x^{(2)}=x^{(1)}-\frac{f(x^{(1)})}{f'(x^{(1)})}$ and so on.
  Thus, for the root $u_2$ starting from the zero order approximation
   $u_2^{(0)}=1$, we have for the function $f(u)$ in (2)
  $f(1)=(\varkappa+1)\delta$. From $f'(u)=-2u+\varkappa\delta+3u^2\delta$ we
  get $f'(1)=-2$. So $u_2^{(1)}=
  1+\frac12(1+\varkappa)\delta$. In the second
  approximation $f(u_2^{(1)})=(\frac{5}{2^2}+\frac{3\varkappa}{2}+
  \frac{\varkappa^2}{2^2})\delta$. As for $f'(u_2^{(1)})$
  we may in the considered approximation still take $f'(1)=-2$.
  So we get $u_2^{(2)}=1+\frac12(1+\varkappa)\delta+
  (\frac{5}{2^3}+\frac{3\varkappa}{2^2}+
  \frac{\varkappa^2}{2^3})\delta^2$. Continuing this
  process, we find
  $$
  u_2=1+\frac12(1+\varkappa)\delta+(\frac{5}{2^3}+\frac{3\varkappa}{2^2}+
  \frac{\varkappa^2}{2^3})\delta^2+(1+\frac{3\varkappa}{2}+
  \frac{\varkappa^2}{2})\delta^3+\cdots.                                         \eqno(5)
  $$
  Now the root $u_1$ can be obtained from here if we note that according to
  (2) $f(u)\equiv f(u,\delta)=f(-u,-\delta)$. Hence,
  $$
  u_1\equiv u_1(\delta)=-u_2(-\delta)=-1+\frac12(1+\varkappa)\delta-
  (\frac{5}{2^3}+\frac{3\varkappa}{2^2}+
  \frac{\varkappa^2}{2^3})\delta^2+(1+\frac{3\varkappa}{2}+
  \frac{\varkappa^2}{2})\delta^3+\cdots.                                          \eqno(6)
  $$

  The expansion for $u_3$ can be obtained in the same way as for $u_2$
  $$
  u_3=\frac{1}{\delta}-(1+\varkappa)\delta-(2+3\varkappa+
  \varkappa^2)\delta^3+\cdots                                                         \eqno(7)
  $$
  It is easy to check that $u_1+u_2+u_3=\delta^{-1}$ as it should be.

  From (2) and (3) we have in agreement with [2-4]
  $$
 \vartheta_{1/2}=\frac{1}{\sqrt{\delta}}
\int_0^{u_2}\frac{du}{\sqrt{(u-u_3)(u-u_2)(u-u_1)}}=
$$
$$
=\frac{2}{\sqrt{(u_3-u_1)\delta}}F(\phi,\kappa),\quad
 \kappa=\left(\frac{u_2-u_1}{u_3-u_1}\right)^{1/2},\quad
 \sin^2 \phi=\frac{1-u_1u_3^{-1}}{1-u_1u_2^{-1}}.                  \eqno(8)
  $$
  Here $F(\phi,\kappa)$ is the elliptical integral.
  Using (5), (6) and (7) we get, retaining terms up to $\delta^2$
  $$
  \kappa^2=2\delta(1-\delta+\cdots),\quad \sin^2 \phi=\frac12\{1+
  \frac{3+\varkappa}{2}\delta+O(\delta^3)\}.                               \eqno(9)
$$
From here
$$
\sin\phi=\frac{1}{\sqrt2}[1+\frac{3+\varkappa}{2^2}\delta-\frac{(3+\varkappa)^2}{2^5}
\delta^2+\cdots], \quad \cos\phi=\frac{1}{\sqrt2}[1-\frac{3+\varkappa}{2^2}\delta+\cdots].                                                      \eqno(10)
$$
To get $\phi$ from the expression for $\sin^2\phi$ it is worth-while to make the substitution
 $\phi=\frac{\pi}{4}+\psi$, $\sin^2\phi=\frac12+\frac12\sin2\psi$.
(This is especially useful when more terms are needed than we retain here, see the next subsection.) So for $\psi$ we have
$$
\sin2\psi=\frac{3+\varkappa}{2}\delta+O(\delta^3). \eqno(11)
$$
So $2\psi=\frac{3+\varkappa}{2}\delta+O(\delta^3)$. Hence,
$$
 \phi=\frac{\pi}{4}+\psi=
\frac{\pi}{4}+
   \frac{3+\varkappa}{2^2}\delta+O(\delta^3).     \eqno(12)
 $$

  As $\delta$ is assumed to be small, $\kappa^2$ is small and we can use
   the expansion, see equation (5) in section (13.6) in [6]
  $$
 F(\phi,\kappa)=S_0+\frac{1}{2}S_2\kappa^2+\frac{3}{2^3}S_4\kappa^5+\cdots,
 \quad S_{2n}\equiv S_{2n}(\phi)=\int_0^{\phi}(\sin t)^{2n}dt;    \eqno(13)
  $$
$$
S_2=\frac12(\phi-\sin\phi\cos\phi);\quad S_4=\frac38\phi-\frac38\sin\phi\cos\phi-\frac14
\sin^3\phi\cos\phi.                                                         \eqno(14)
$$
From (6) and (7) we have
$$
(u_3-u_1)\delta=1+\delta-\frac32(1+\varkappa)\delta^2+\cdots.                     \eqno(15)
$$
Using equation (3.6.19) in [7] we obtain from here
$$
[(u_3-u_1)\delta]^{-1/2}=1-\frac12\delta+\left(\frac38+\frac{\varkappa}{4} \right)3\delta^2+\cdots.                                                       \eqno(16)
$$
Performing remaining calculation, we get for $\vartheta_{1/2}$ in (8)
$$
\vartheta_{1/2}=\frac{\pi}{2}+(1+\frac{\varkappa}{2})\delta+
\pi\left(\frac{15}{32}+\frac{3\varkappa}{8}\right)\delta^2+\cdots.                  \eqno(17)
$$

The scattering angle is $\theta=2\vartheta_{1/2}-\pi$. So
$$
\frac{\theta}{2}=\vartheta_{1/2}-\frac{\pi}{2}=(1+\frac{\varkappa}{2})\delta+
\pi\left(\frac{15}{32}+\frac{3\varkappa}{8}\right)\delta^2+\cdots.                  \eqno(18)
$$
Using  equation (3.6.25) in [7] we find
$$
\delta=\frac{\theta}{2+\varkappa}-\frac{5+4\varkappa}{(2+\varkappa)^3}\frac{3\pi}{2^4}\theta^2+
\cdots.                                                              \eqno(19)
$$
From here with the help of (3.6.25) in [7] we find
$$
\delta^{-2}=\left(1+\frac{\varkappa}{2}\right)^2\left(\frac{2}{\theta}\right)^2\{1+\frac{5+
4\varkappa}{(2+\varkappa)^2}\frac{3\pi}{8}\theta+\cdots \}.                  \eqno(20)
$$
This gives the classical integral cross section
$$
\sigma_{cl}(\theta)=\pi\rho^2(\theta),\quad \delta^{-2}\equiv\rho^2/r_{g}^2. \eqno(21)
$$

If we compare (18) with correspondind equation (10.14) in [4] we find that only the
 leading term is the same. The preceding equation (10.13) in [4] contains errors in
powers of the root of the cubic.

The differential cross section is obtained from (21) by differentiation, multiplying by $\pi$ and changing the
averall sign in the right hand side:
$$
d\sigma_{cl}(\theta)=2\pi r_g^2(1+\frac12\varkappa)^2\frac{1}{y^3}\{1+
\frac{5+4\varkappa}{(2+\varkappa)^2}\frac{3\pi}{8}y+\cdots\}dy,\quad y=\frac{\theta}{2}.                                                                                      \eqno(21a)
$$

In quantum picture the first Born approximation for scalar particle cross section is given in equation (39) below. It contains only even powers of $y$ as in Coulomb scattering [1]. This
should mean that corrections to the first Born approximation must contain classical terms
independent of $\hbar$. In Coulomb scattering these classical terms in higher approximations
may be expected only at unrealistically high $\alpha Z>>1$.\\
{\bf{B) Massless particle}}\\
In this subsection we obtain more terms of the expansions for the case of massless particle
i.e. for $\varkappa=0$, see (2).
Continuing the process described in getting (5) we find
$$
u_2=1+\frac12\delta+\frac{5}{2^3}\delta^2+\delta^3+\frac{3\cdot7\cdot11}{2^7}\delta^4+
\frac{7}{2}\delta^5+\cdots,      \eqno(22)
$$
$$
u_3=\frac{1}{\delta}-\delta-2\delta^3-7\delta^5+\cdots,     \eqno(23)
$$
$$
  u_1\equiv u_1(\delta)=-u_2(-\delta)=-1+\frac12\delta-
  \frac{5}{2^3}\delta^2+
  \delta^3-\frac{231}{2^7}\delta^4+\frac{7}{2}\delta^5+\cdots.                                                                                                            \eqno(24)
$$

Using these expressions we find for $\kappa^2$ and $\sin^2\phi$, see (7)
$$
\kappa^2=2\delta\{1-\delta+\frac{5^2}{2^3}\delta^2-\frac{3\cdot7}{2^2}\delta^3-
\frac{7\cdot281}{2^7}\delta^4+\cdots\},           \eqno(25)
$$
$$
\sin^2\phi=\frac12\{1+\frac32\delta+\frac{3\cdot11}{2^4}\delta^3
+\frac{3^2\cdot5\cdot37}{2^8}\delta^5+\cdots\}.                            \eqno(26)
$$
From here
$$
\sin\phi=\frac{1}{\sqrt2}\{1+\frac34\delta-\frac{3^2}{2^5}\delta^2+\frac{3\cdot53}{2^7}\delta^3
-\frac{3^2\cdot221}{2^{11}}\delta^4+\frac{3^2\cdot3941}{2^{13}}\delta^5+\cdots\}. \eqno(27)
$$
The term with $\delta^5$ is needed only if we want to obtain $\phi$ from (27). But we proceed as in obtaining (11):
$$
\sin2\psi=\frac32\delta\{1+\frac{11}{2^3}\delta^2+\frac{3\cdot5\cdot37}{2^7}\delta^4+\cdots\}.
                                                                          \eqno(28)
$$
From here
$$
2\psi=\frac{3}{2}\delta+\frac{3\cdot7}{2^3}\delta^3+\frac{3^2\cdot167}{2^5\cdot5}\delta^5+
\cdots.                                                                     \eqno(29)
$$
Hence
$$
\phi=\frac{\pi}{4}+\psi=\frac{\pi}{4}+\frac{3}{2^2}\delta+\frac{3\cdot7}{2^4}\delta^3+
\frac{3^2\cdot167}{2^6\cdot5}\delta^5+
\cdots.                                                                       \eqno(30)
$$
From (27) we obtain
$$
\cos\phi=\frac{1}{\sqrt2}\{1-\frac34\delta-\frac{3^2}{2^5}\delta^2-
\frac{3\cdot53}{2^7}\delta^3
-\frac{3^2\cdot221}{2^{11}}\delta^4+\cdots\}. \eqno(31)
$$
In equation (16) (with $\varkappa=0$) we have now more terms:
$$
[(u_3-u_1)\delta]^{-1/2}=1-\frac12\delta+\frac{3^2}{2^3}\delta^2-\frac{7}{2^2}\delta^3+
\frac{5^2\cdot23}{2^7}\delta^4-\frac{3^3\cdot5}{2^4}\delta^5+\cdots.       \eqno(32)
$$
 Using expressions for $\sin\phi$ and $\cos\phi$ in (27) and (31) we obtain $S_{2n}$ which
are polynomials in $\sin\phi$ and $\cos\phi$. Then using also (25) we find similarly to (13)
$$
 F(\phi,\kappa)=S_0+\frac{1}{2}S_2\kappa^2+\frac{3}{2^3}S_4\kappa^4+
\frac{5}{2^4}S_6\kappa^6+\frac{5\cdot7}{2^7}S_8\kappa^8+
\frac{3^2\cdot7}{2^8}S_{10}\kappa^{10}+ \cdots=\frac{\pi}{4}+\left(\frac{\pi}{2^3}+\frac12  \right)\delta                                                            \eqno(33)
$$
$$
+\left(\frac{\pi}{2^6}+\frac{1}{2^2}\right)\delta^2+
\left(\frac{39\pi}{2^7}+\frac{43}{2^4\cdot3}\right)\delta^3+
\left(\frac{313\pi}{2^{12}}+\frac{5^2}{2^3\cdot3}\right)\delta^4+
\left(\frac{7\cdot1487\pi}{2^{13}}+\frac{12689}{2^8\cdot3\cdot5}\right)\delta^5. \eqno(34)
$$
Using (32) we get from (8) and (34)
$$
\vartheta_{1/2}=\frac{\pi}{2}+\delta+
\frac{3\cdot5\pi}{2^5}\delta^2+\frac{2^3}{3}\delta^3+
\frac{3^2\cdot5\cdot7\cdot11\pi}{2^{11}}\delta^4+\frac{2^3\cdot7}{5}\delta^5  \cdots,                                                                                         \eqno(35)
$$
and similarly to (19) we obtain
$$
\delta=y\{1-\frac{3\cdot5\pi}{2^5}y+\left(\frac{3^2\cdot5^2\pi^2}{2^9}-\frac{2^3}{3}\right)y^2  +\left(\frac{5\cdot1867\pi}{2^{11}}-\frac{3^3\cdot5^4\pi^3}{2^{15}}\right)y^3+
$$
$$
\left(\frac{2^3\cdot19}{3\cdot5}- \frac{3^2\cdot5^2\cdot7\cdot157\pi^2}{2^{15}}+\frac{3^4\cdot5^4\cdot7\pi^4}{2^{19}}\right)y^4+
\cdots\},\quad y=\frac{\theta}{2}=\vartheta_{1/2}-\frac{\pi}{2}.          \eqno(36)
$$
As in (20) we find
$$
\delta^{-2}\equiv\frac{\rho^2}{r_g^2}=y^{-2}\{1+c_1y+c_2y^2+c_3y^3+c_4y^4+\cdots\},\eqno(37)
$$
$$
\quad c_1=\frac{3\cdot5\pi}{2^4},\quad c_2=\frac{2^4}{3}-\frac{3^2\cdot5^2\pi^2}{2^{10}}, \quad
$$
$$
c_3=-\frac{5\cdot331\pi}{2^{10}}+\frac{3^3\cdot5^3\pi^3}{2^{14}} \quad
c_4=\frac{2^4}{3\cdot5}+\frac{3^2\cdot5^2\cdot331\pi^2}{2^{15}}-
\frac{3^4\cdot5^5\pi^4}{2^{20}}.                     \eqno(37a)
$$
The approximate value for $c_i, i=1, 2, 3, 4$ are
$$
c_1=2.945,\quad c_2=3.165,\quad c_3=1.310,\quad c_4=-0.016.   \eqno(37b)
$$
The differential cross section is obtained from (37) by differentiation
$$
d\sigma(\theta)_{cl}=\pi r_g^2\frac{2^3}{\theta^3}\{1+c_1\frac{\theta}{2^2}-c_3\frac{\theta^3}{2^4}-
c_4\frac{\theta^4}{2^4}+\cdots\}d\theta. \eqno(38)
$$

It is interesting to compare this classical cross section with the quantum one. For
 massless scalar particle we have in the first Born approximation [8]
$$
d\sigma_{sc}=2\pi r^2_{g}(1+\frac{\varkappa}{2})^2\frac{\cos y}{\sin^3y}dy=2\pi r^2_{g}(1+\frac{\varkappa}{2})^2\frac{1}{y^3}\{1-
\frac{y^4}{15}+\cdots\}dy, \quad y=\frac{\theta}{2}.   \eqno(39)
$$
The substitution $y\to-y$ changes the sign of $\sin y$ and $y$. Hence the expression in
braces in (39) can contain only even powers of $y$. We note also that $c_4$ in (38) is
negative ($c_4=-0.016$) in contrast with corresponding coefficient $1/15$ in (39).

The quantum cross section for photon contains an additional factor:
$$
d\sigma_{\gamma}=d\sigma_{sc}\cos^4y,  \eqno(40)
$$
see[8] and references therein. The factor $d\sigma_{sc}$ in (40) and below is taken
at $\varkappa=0$ i.e. for massless particle.
So for $ y>0$ we have
$$
d\sigma_{\gamma}<d\sigma_{sc}
$$

For graviton [9], [10]:
$$
d\sigma_{g}=d\sigma_{sc}\frac18(1+6\cos^\theta+\cos^4\theta)=d\sigma_{sc}(\cos^8y+\sin^8y).                                                                                \eqno(40a)
$$
Hence $d\sigma_{g}<d\sigma_{\gamma}$ for $0<y<<1$. So for small angle scattering the
cross section is smaller for particle with higher spin. It seems reasonable to expect from
these facts that spin facilitates forward scattering of a particle described by a wave
packet.
\section{Small angle classical scattering by the linear approximation of the Schwarzschild field}

To see the effects of the nonlinearity of the Schwarzschild field on scattering, we consider
in this Section the linear approximation of the isotropic coordinates. This corresponds to the quantum treatment of scattering in [8-10]. Only massless particle is considered here.

Proceeding in the same way as in [3] and [4] we obtain instead of (2)
$$
  \frac{du}{d\vartheta}=\pm\sqrt{\frac{f(u)}{1-u\delta}},\quad f(u)=1-u^2+(u^3+u)\delta=
\delta(u-u_1)(u-u_2)(u-u_3).  \eqno(41)
 $$
So we use in this Section the same notation as before , but with a somewhat different meaning. For $\vartheta_{1/2}$ we get a more complicated expression:
$$
  \vartheta_{1/2}=\frac{1}{\sqrt{\delta}}\int_0^{u_2}
\sqrt{\frac{1-u\delta}{(u_3-u)(u_2-u)(u-u_1)}}du=\frac{1}{\sqrt{u_3\delta}}
\int_0^{u_2}\sqrt{\frac{1-u\delta}{(1-uu^{-1}_3)}}\frac{du}{\sqrt{R(u)}}.   \eqno(42)
$$
Here $R(u)=(u_2-u)(u-u_1)$.

From (41) we find instead of (22-24)
$$
u_2=1+\delta+\frac{3}{2}\delta^2+3\delta^3+\frac{5\cdot11}{2^3}\delta^4+
17\delta^5+\cdots,      \eqno(43)
$$
$$
u_3=\frac{1}{\delta}-2\delta-6\delta^3-34\delta^5+\cdots,     \eqno(44)
$$
$$
  u_1\equiv u_1(\delta)=-u_2(-\delta)=-1+\delta-
  \frac{3}{2}\delta^2+
  3\delta^3-\frac{5\cdot11}{2^3}\delta^4+17\delta^5+\cdots.                                                                                                            \eqno(45)
$$

For $\delta<<1$ we have $u_3>>1$, Hence $u/u_3<<1$ in the r.h.s. of(42).
So to evaluate (42) we can proceed as follows
$$
\frac{1-u\delta}{1-uu^{-1}_3}=\frac{(1-u\delta-2u\delta^3-10u\delta^5)+2u\delta^3
(1+5\delta^2)}
{1-u\delta(1+2\delta^2+10\delta^4
\cdots)}=
$$
$$
1+\frac{2u\delta^3(1+5\delta^2)}{1-u\delta-2u\delta^3+\cdots}=1+2u\delta^3[1+u\delta+
(u^2+5)\delta^2+\cdots].                                                            \eqno(46)
$$
From here
$$
\left(\frac{1-u\delta}{1-uu^{-1}_3}\right)^{1/2}=1+u(\delta^3+5\delta^5)+
u^2\delta^2+u^3\delta^5+\cdots.                                           \eqno(47)
$$
Using this in (42) we have
$$
\vartheta_{1/2}=\frac{1}{\sqrt{u_3\delta}}\{I_0+(\delta^3+5\delta^5)I_1+\delta^4I_2
+\delta^5I_3+\cdots  \},\quad I_n=\int_0^{u_2}\frac{u^ndu}{\sqrt{R(u)}}.  \eqno(48)
$$
Evaluating these integrals we find
$$
I_0=\frac{\pi}{2}+\arcsin{\frac{u_2+u_1}{u_2-u_1}},\quad I_1=
\sqrt{-u_2u_1}+\frac{u_1+u_2}{2}I_0,
$$
$$ I_2=3\frac{u_1+u_2}{4}\sqrt{-u_1u_2}+\{\frac{3(u_1+u_2)^2}{8}-\frac{u_1u_2}{2}  \}I_0,
 R(0)=-u_1u_2, \quad R(u_2)=0.                                                                              \eqno(49)
$$
$R(u)$ is defined below (42).
As for $I_3$ we need it only at $\delta=0$. Then its value is $2/3$.
Using(43-45) we get
$$
I_0=\frac{\pi}{2}+\delta+\frac{5}{3}\delta^3+\frac{87}{10}\delta^5+\cdots,\quad I_1=1+\frac{\pi}{2}\delta+2\delta^2+\cdots,\quad I_2=\frac{\pi}{4}+2\delta+\cdots  \eqno(50)
$$
$$
\frac{1}{\sqrt{u_3\delta}}=1+\delta^2+\frac{9}{2}\delta^4+O(\delta^6).
$$
Now for (42) we obtain
$$
\vartheta_{1/2}=\frac{\pi}{2}+\delta+
\frac{\pi}{2}\delta^2+\frac{11}{3}\delta^3+
3\pi\delta^4+\frac{383}{15}\delta^5  \cdots,                                                                                         \eqno(51)
$$
(We have used this method to check the equation (35))
From here
$$
\delta=y\{1-\frac{\pi}{2}y+\left(\frac{\pi^2}{2}-\frac{11}{3}\right)y^2  +\left(\frac{37\pi}{6}-\frac{5\pi^3}{8}\right)y^3+
$$
$$
\left(\frac{74}{5}- \frac{41\pi^2}{4}+\frac{7\pi^4}{2^3}\right)y^4+
\cdots\},\quad y=\frac{\theta}{2}=\vartheta_{1/2}-\frac{\pi}{2}.          \eqno(52)
$$

As in (20) we obtain $\delta^{-2}$ in the form (37) where now
$$
c_1=\pi,\quad c_2=\frac{22}{3}-\frac{\pi^2}{4}, \quad
c_3=-\frac{4\pi}{3}+\frac{\pi^3}{4}, \quad
c_4=\frac{161}{15}+2\pi^2-
\frac{5\pi^4}{16}.                     \eqno(53)
$$
With these $c_i$ the equation (38) holds.
Here the approximate value  of $c_i$ are
$$
c_1=3.1416,\quad c_2=4.8660,\quad c_3=3.5627,\quad c_4=0.0322.   \eqno(53a)
$$
Comparing with (37b) we see how the difference between $c_i$ there and here increases
with $i$.

\section{Small angle classical scattering by an interval of gravitational field}

This problem is usually avoided because all coordinate systems are equivalent in general
relativity. Here we assume that in the privileged system the observed deflection is
given by the tangent to the trajectory. There are some reason to think that the
privileged system must be isotropic one [11] and we take it from general relativity.

For tangent we have
$$
\tan\varphi=\frac{dy}{d\vartheta}/\frac{dx}{d\vartheta}=
\frac{\frac{dr}{d\vartheta}\sin\vartheta+r\cos\vartheta}
{\frac{dr}{d\vartheta}\cos\vartheta-r\sin\vartheta},\quad x=r\cos\vartheta,\quad y=r\sin\vartheta, \eqno(54)
$$
or in terms of $u=\rho/r$:
$$
\tan\varphi=\frac{\frac{du}{d\vartheta}\sin\vartheta-u\cos\vartheta}
{\frac{du}{d\vartheta}\cos\vartheta+u\sin\vartheta}. \eqno(55)
$$

So we have to know $u(\vartheta)$ or $\vartheta(u)$. For a massless particle in the isotropic system we have
$$
\frac{du}{d\vartheta}=\pm\sqrt{\frac{(1+\frac14u\delta)^6}{(1-\frac14u\delta)^2}-u^2}.\eqno(56)
$$
To get $u(\vartheta)$ from here is a mach more difficult job than in the case of the
standard Schwarzschild system utilized in (2).

In the following we assume for simplicity that $\delta<<1$ and $1-u^2$ is of order of unity.
(The latter assumption is not needed for the final result (60)) Then for the ingoing half of the trajectory we may write
$$
\frac{du}{d\vartheta}=\sqrt{1-u^2}\left(1+\frac{u\delta}{1-u^2}+O(\delta^2)\right). \eqno(57)
$$
From here we get
$$
\vartheta=\arcsin u+\Delta,\quad \Delta=\left(1-\frac{1}{\sqrt{1-u^2}}\right)\delta.  \eqno(58)
$$
So
$$
\sin\vartheta=u+\Delta\sqrt{1-u^2}=u+(\sqrt{1+u^2}-1)\delta,\quad \cos\vartheta=\sqrt{1-u^2}-\Delta u. \eqno(59)
$$
Then, using (57) and (59), we find from (55)
$$
\tan\varphi\approx \varphi=(1-\sqrt{1-u^2})\delta=(\frac12u^2+\frac18u^4+\cdots)\delta
                                                                        \eqno(60)
$$
This is the contribution from the part of the trajectory beginning at $u=0$ and ending
at $u$.

We note here that from (59) we can obtain the trajectory in the form
$$
u=\sin\theta+(1-\cos\theta)\delta,    \eqno(57a)
$$
which is equation (40.6) in \$ 40 in [5].
For this in the term with $\delta$ in the second equation in (59)  we replace
$\sqrt{1-u^2}$ by its zero order value $\cos\theta$.

As is well known the leading term of the small angle classical deflection can be obtained by simple
mechanical considerations, see \S 39, Problem 2 in [12], or equation (4.41) in Ch2 in [13].
In our case the contribution from the whole trajectory is
$$
\varphi=r_g\rho\int_{-\infty}^{\infty}\frac{dx}{(x^2+\rho^2)^{3/2}}=2\delta.  \eqno(61)
$$
It is assumed here that on the r.h.s. the trajectory may be taken as a straight line; $y\cos\vartheta=\rho.$
The contribution from the same part of the trajectory as in (60) is ($x=r\cos\vartheta; y=
r\sin\vartheta)$
$$
r_g\rho\int_{x}^{\infty}\frac{dx}{(x^2+\rho^2)^{3/2}}=\frac{r_g}{\rho}\left(1-\frac{x\rho^{-1}}{\sqrt{1+_(x\rho^{-1})^2}}
\right).                                                                           \eqno(62)
$$
As
$$
\frac{x}{\rho}=\frac{r}{\rho}\cos\vartheta=\frac{1}{u}\sqrt{1-u^2}
$$
the r.h.s. of (62) is equal to that of (60).
Only smallness of $\varphi$ is used in obtaining (62). If $u<<1$ then $\delta$  may be of order unity. If $\delta<<1$ then $u$ may be of order unity.

On the surface of the Sun $0\leq\varphi\leq\delta$. Zero corresponds to the radial trajectory
 $\rho=0$, $\delta$ corresponds to the trajectory touching the surface. For $u<<1$ we have from (62) $\varphi=(r_g\rho)/(2r^2)$.

If we use the standard Schwarzschild coordinate system, see (1) and (2), we get instead of
$\sin\vartheta$ in (59)
$$
\sin\vartheta=u+[\sqrt{1+u^2}+\frac{u^2}{2}-1]\delta
$$
and
$$
\tan\varphi\approx \varphi=(1-\sqrt{1-u^2}(1+\frac{u^2}{2}))\delta=\frac18u^4\delta+\cdots.
                                                                        \eqno(63)
$$
instead of (60). (As it should be, for $u=1$
 both (60) and (63) give $\tan\varphi=\vartheta_{1/2}=\delta$.) So the measurements of $\varphi$ in the vicinity of the Sun can decide
which coordinate system is the privileged one.

Finally, the contribution to $\varphi$ from a finite interval of trajectory from $x$ to $\tilde x$ is
$$
r_g\rho\int_{x}^{\tilde x}\frac{dx'}{(x'^2+\rho^2)^{3/2}}=\frac{r_g}{\rho}
[\frac{\tilde x}{\sqrt{\rho^2+\tilde x^2}}-\frac{x}{\sqrt{\rho^2+x^2}}].  \eqno(64)
$$
see (61). In (64) $x$ and $\tilde x$ may lie on the ingoing $x<0$ and outgoing $\tilde ><0$
halves of the trajectory respectively. In more precise approach (in terms of $u$) we
must treat each half of the trajectory separately.
\section*{Acknowledgements}
I am indebted to V.I. Ritus for stimulating dicussions. The work was supported by Scientific Schools and Russian Fund for Fundamental Research (Grants 4401.2006.2and 08-02-01118).
\section*{References}
1. A.I.Nikishov, hep-th:0804.2149.\\
2. Y.Hagihara, Japan. J. Astron. Geophys. {\bf 8}, 67 (1931).\\
3. C.G. Darwin, Proc. Roy. Soc. London. A249, 180 (1959).\\
4. B.Mielnik, J. Plebanski, Acta Phys. Polon, {\bf21}, 239 (1962).\\
5.  C.W.Misner, K.S.Thorne, J.A.Wheeler, {\sl Gravitation.} San
   Francisco(1973).\\
6. H.Bateman, A. Erdelyi,{\sl Higher transcendental function}, Vol.3 (McGuire-Hill, New York,      1953).\\
7. M. Abramowitz and I. Stegun. { \sl Handbook of Mathematical Functions},
  (National Bureau of Standards, 1964).\\
8. N.V.Mitskevish, Zh. Eksper. Teoret. Fiz.,{\bf{34}},1656,(1958).\\
9. D.J. Gross and R. Jackiw, Phys. Rev.,{\bf{166}}, 1287 (1966).\\
10  F.A. Berends, R. Gastmans, Nucl. Phys., {\bf{B88}}, 99 (1975).\\
11. A.I. Nikishov, gr-qc:0710.4445. \\
12. L.D. Landau and E.M. Lifshitz,The Classical Theory of Fields (Reading, Mass 1952).\\
13. J. Schwinger, {\sl Particles, Sources, and Fields} (Addison-Wesley 1970).

\end{document}